\documentclass[%
reprint,
 amsmath,amssymb,
 aps,
]{revtex4-1}

\usepackage{graphicx}
\usepackage{dcolumn}
\usepackage{bm}

\begin{document}


\title{Direct Evidence for Suppression of the Kondo Effect due to Pure Spin Current}

\author{K. Hamaya}
 \email{hamaya@ee.es.osaka-u.ac.jp}
\author{T. Kurokawa}
\author{S. Oki}
\author{S. Yamada}
\author{T. Kanashima}
\affiliation{Graduate School of Engineering Science, Osaka University,Toyonaka 560-8531, Japan}
\author{T. Taniyama}%
 \email{taniyama.t.aa@m.titech.ac.jp}
\affiliation{Materials and Structures Laboratory, Tokyo Institute of Technology, 4259 Nagatsuta, Midori-ku, Yokohama 226-8503, Japan}%

\date{\today}

\begin{abstract}
We study the effect of a pure spin current on the Kondo singlet in a diluted magnetic alloy using non-local lateral spin valve structures with highly spin polarized Co$_{2}$FeSi electrodes. Temperature dependence of the non-local spin signals shows a sharp reduction with decreasing temperature, followed by a plateau corresponding to the low temperature Fermi liquid regime below the Kondo temperature ({\it T}$_{\rm K}$). The spin diffusion length of the Kondo alloy is found to increase with the evolution of spin accumulation. The results are in agreement with the intuitive description that the Kondo singlet cannot survive any more in sufficiently large spin accumulation even below {\it T}$_{\rm K}$. 
\end{abstract}

\maketitle

According to the Anderson model\cite{Yoshida}, the $s$-$d$ mixing or hybridization of a localized impurity spin and the surrounding conduction electrons results in the formation of a spin singlet state in diluted magnetic alloys\cite{Kondo,Franck,Applebaum,Loram,Monod,Haldane} or artificial nanostructures such as a quantum dot (QD) \cite{Gold,Cronenwett,Gold2,Wiel}. This many-body effect, so-called Kondo effect, is currently being researched extensively due to its rich physics in condensed matter \cite{Sato,Lee,Buizert,Yamaguchi,Bultelaar,OBrien,Batley,Martinek1,Martinek2,Choi,Barnas,Ralph,Gossard,Hamaya,Hauptmann,Heersche,Taniyama,Kobayashi}; a logarithmic increase in the resistivity of diluted magnetic alloys below the characteristic temperature, i.e., the Kondo temperature ({\it T}$_{\rm K}$), is a representative aspect of the Kondo effect. Among them, the effect of spin polarized electrons on the Kondo singlet is now receiving great interest from both theoretical \cite{Martinek1,Martinek2,Choi,Barnas} and experimental \cite{Ralph,Hamaya,Heersche,Hauptmann,Taniyama} points of view.  
As illustrated in Fig. 1(a), the low temperature Fermi liquid Kondo regime at {\it T} $<$ {\it T}$_{\rm K}$ is a direct consequence of the spin-flip scattering between the impurity spin and conduction electron spins in a nonmagnetic host in zero magnetic field ($\Delta \epsilon_{\rm d}=\epsilon_{\rm d\uparrow}-\epsilon_{\rm d\downarrow}=0$). 
\begin{figure}[b]
\begin{center}
\includegraphics[width=8cm]{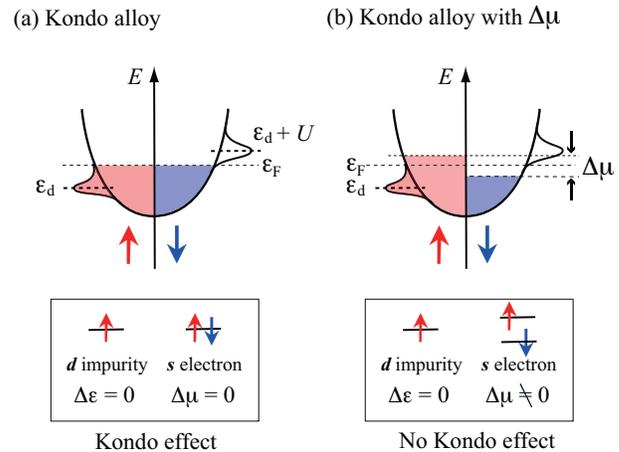}
\caption{(Color online) Schematic illustrations of the density of states (DOS) in a Kondo alloy based on the Anderson model in {\it T} $<$ {\it T}$_{\rm K}$ (a) without and (b) with spin accumulation ($\Delta\mu$) in the host nonmagnetic metal. }
\end{center}
\end{figure}  
If spin accumulation ($\Delta\mu=\mu_{\uparrow}-\mu_{\downarrow}$) defined by the difference in the chemical potentials of spin-up ($\mu_{\uparrow}$) and spin-down ($\mu_{\downarrow}$) electrons is generated in the host nonmagnetic metal\cite{MJ,vanSon}, the spin accumulation has an influence on the Kondo effect and even suppresses the formation of the Kondo singlet due to an additional energy cost for the spin-flip scattering (Fig. 1(b)). Similarly, as strong local exchange interaction\cite{Ralph,Hamaya,Hauptmann} or RKKY interaction\cite{Heersche} manifests itself in the QD coupled with ferromagnetic leads, the Kondo assisted tunneling is suppressed in the Kondo regime, leading to the splitting of the Kondo zero-bias anomaly even in zero magnetic field\cite{Martinek1,Martinek2,Choi,Barnas}.
Although such suppression of the Kondo effect has been observed by injecting spin polarized charge currents\cite{Taniyama}, the effect of a pure spin current, which is a flow of spin angular moment without a net charge current\cite{Jedema,Kimura}, has not yet been elucidated in the Kondo systems because of the difficulty in detecting pure spin current transport in Kondo alloys or QDs.  

In this Letter, we report on direct evidence for the effect of a pure spin current in diluted magnetic Cu(Fe) Kondo alloy (the average Fe concentration of 100 $\sim$ 200 p.p.m)\cite{Franck,Kondo,Applebaum,Monod,Loram}, by using non-local lateral spin valve (LSV) devices with highly spin polarized Co$_{2}$FeSi (CFS) electrodes. The half-metallic characteristics of CFS surely enable us to detect spin signals in the diluted magnetic Kondo alloy even at a very low signal level\cite{Wurmehl,Bombor}. We find that the spin signal arising from pure spin current transport is strongly suppressed below the onset temperature ({\it T} $>$ {\it T}$_{\rm K}$) of $s$-$d$ spin-flip scattering while the spin signal does not decrease any more at {\it T} $<$ {\it T}$_{\rm K}$ because of the collective screening of the magnetic Fe impurity spins by Kondo clouds. The {\it T}$_{\rm K}$ estimated from the pure spin current measurement is in good agreement with that obtained from the temperature dependence of the resistivity. We also demonstrate that the Kondo effect can be tuned by generating larger spin accumulation in the diluted magnetic alloy.

25-nm-thick CFS layers were grown on non-doped FZ-Si(111) substrates by molecular beam epitaxy (MBE) at 60$^{\circ}$C, where Co, Fe and Si were co-evaporated from Knudsen cells\cite{Hamaya1,Yamada}. In-situ reflection high energy electron diffraction patterns of CFS layers clearly exhibited symmetrical streaks, indicating good two-dimensional epitaxial growth\cite{Yamada}. The CFS layers were patterned into the submicron-sized electrodes by using conventional electron beam lithography and Ar ion milling for measurements of non-local spin signals\cite{Kimura1,Hamaya2}. To fabricate the LSVs, 75-nm-thick Cu(Fe) wires bridging the two CFS electrodes with various center-to-center distances ($d$) were patterned by a conventional lift-off technique, together with bonding pads. The interface resistance between CFS and Cu(Fe) is negligible ($\leq$0.1 f$\Omega$m$^{2}$) by carefully cleaning the surface of the CFS layers with low-energy accelerated Ar ion milling\cite{Hamaya2}.  Figure 2(a) shows a scanning electron microscopy image of a fabricated LSV. By using the cross terminal configuration, non-local spin signal measurements were carried out using a conventional current-bias lock-in technique (173 Hz) at various temperatures.
\begin{figure}[b]
\begin{center}
\includegraphics[width=8.0cm]{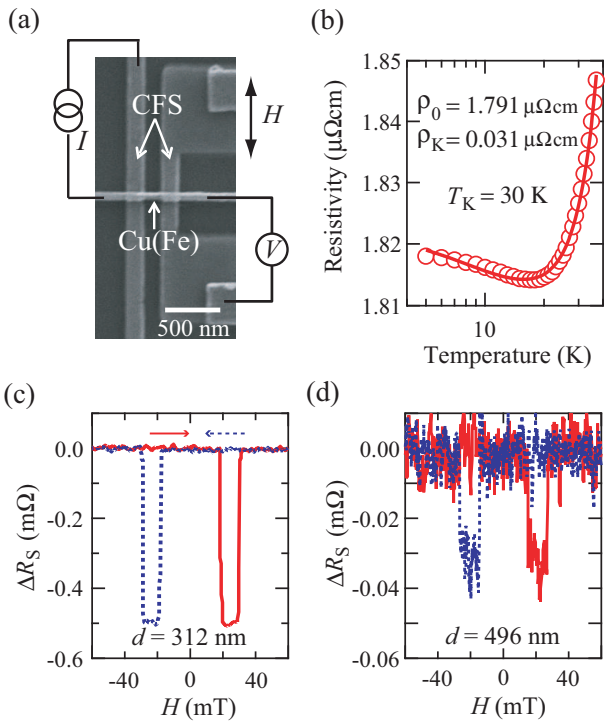}
\caption{(Color online) (a) A scanning electron micrograph of a CFS-Cu(Fe)-CFS LSV. (b) Temperature dependent resistivity of Cu(Fe) used in this study. Non-local spin signals of CFS-Cu(Fe)-CFS LSVs with (c) $d =$ 312 nm and (d) $d =$ 496 nm, respectively, measured at 5 K with $I =$ 1.0 mA. }
\end{center}
\end{figure}  

We confirmed the Kondo effect in the Cu(Fe) wire of the LSV by measuring the temperature dependent resistivity, as shown in Fig. 2(b). The resistivity minimum ({\it T}$_{\rm min}$) can be seen at around 20 K, and the resistivity increases with decreasing temperature. An empirical functional model for the Kondo resistivity can be expressed as $\rho(T) = \rho_{0}$ + $AT^{2}$ + $BT^{5}$ + $\rho_{K}$/$\left[1+(2^{1/s} - 1)(T/T_{K})^{2}\right]^{s}$\cite{Gold2,OBrien}, where $\rho_{0}$ is the residual resistivity and the second and third terms are the electron-electron and electron-phonon scattering contributions, respectively. The last term arises from the Kondo effect\cite{Gold2,OBrien}. Assuming $s \sim$ 0.22 in a spin-(1/2) Kondo system\cite{Gold2}, a best fit curve with the above function was obtained as shown in the solid curve in Fig. 2(b), yielding {\it T}$_{\rm K} =$ 30 K. The values of $\rho_{\rm 0}$ and $\rho_{\rm K}$ are 1.791 $\mu\Omega$cm and 0.031 $\mu\Omega$cm, respectively. These analyses ensure that the Cu(Fe) wire we used shows a typical Kondo effect with {\it T}$_{\rm K} =$ 30 K\cite{Kondo,Applebaum,Monod,Loram,Haldane}.

Figures 2(c) and 2(d) show typical non-local spin signals ($\Delta R_{\rm S}$) detected by using CFS-Cu(Fe)-CFS LSVs (Fig. 2(a)) with $d =$ 312 and 496 nm, respectively, at 5 K, where $\Delta R_{\rm S}$ is calculated by $\Delta V$/$I$. Clear hysteretic spin signals are observed even for the spin injection into Cu(Fe) Kondo alloys.  
This is the first experimental observation of the lateral transport of a pure spin current in a Kondo alloy.
It should be noted that the spin signals, i.e., $\Delta R_{\rm S}$, was one order or two orders of magnitude smaller than those observed in typical Cu-based LSVs with a CFS spin injector and a detector in our previous work\cite{Kimura1,Hamaya2}. This is caused by the small spin diffusion length of Cu(Fe) ($\lambda_{\rm Cu(Fe)}$)\cite{Taniyama}. Using the CFS-Cu(Fe)-CFS LSV devices with various $d$ values, we also measured temperature dependent $\Delta R_{\rm S}$ at $I =$ 1.0 mA as shown in Fig. 3(a). 
\begin{figure}[t]
\begin{center}
\includegraphics[width=8.0cm]{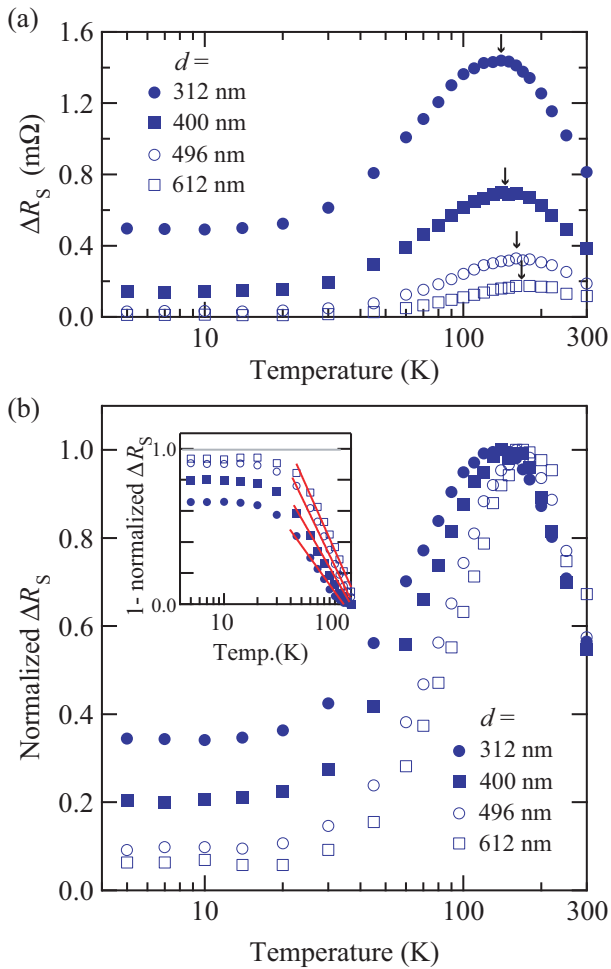}
\caption{(Color online) (a) Temperature dependence of $\Delta R_{\rm S}$ for CFS-Cu(Fe)-CFS LSVs with various $d$ for $I =$ 1.0 mA. (b) Normalized $\Delta R_{\rm S}$ as a function of temperature for the various LSVs. The inset shows a plot of (1-normalized $\Delta R_{\rm S}$) versus temperature. The red lines are fits by using a function representing the logarithmic temperature variation. }
\end{center}
\end{figure}  
For all the LSVs, there is a clear maximum of $\Delta R_{\rm S}$ at a characteristic temperature which we refer to as {\it T}$_{\rm max}$. {\it T}$_{\rm max}$ is different for each LSV (see arrows). 
We find that the $\Delta R_{\rm S}$ values at 5 K are smaller than those at 300 K for all the LSVs. This feature is largely different from those observed in conventional Cu-based LSVs with CFS\cite{Hamaya2}. 
Of particular significance here is that the $\Delta R_{\rm S}$ is markedly decreased as the temperature is lowered from {\it T}$_{\rm max}$, and finally saturates below $\sim$ 20 K. Because spin signal $\Delta R_{\rm S}$ for a fixed terminal distance $d$ provides a measure of spin accumulation, the reduction in $\Delta R_{\rm S}$ at low temperatures is a signature of the enhanced spin-flip scattering events.

In order to compare the temperature variation in the spin-flip scattering events for the LSVs with the different $d$ values, the normalized $\Delta R_{\rm S}$ values are also shown in Fig. 3(b) as a function of temperature, where the normalized $\Delta R_{\rm S}$ is defined as $\Delta R_{\rm S}$ divided by the maximum $\Delta R_{\rm S}$. Note that the normalized $\Delta R_{\rm S}$ values at {\it T} $\le$  {\it T}$_{\rm max}$ show significant $d$ dependence, in striking contrast to no $d$ dependence in the temperature region {\it T} $\ge$ {\it T}$_{\rm max}$. The large $d$ dependence of the normalized $\Delta R_{\rm S}$ at {\it T} $\le$  {\it T}$_{\rm max}$ indicates that the spin-flip scattering strongly depends on spin accumulation and it is intensively suppressed at a small $d$ due to the large spin accumulation, consistent with the Kondo effect as described before. We also define the value, ($1 -$ normalized $\Delta R_{\rm S}$), at {\it T} $\le$ {\it T}$_{\rm max}$, as shown in the inset of Fig. 3(b). The red line shows a fit with the logarithmic function, $a$[1-$b$log($\frac{{\it T}}{{\it T}_{\rm K}}$)], in the temperature range 40 K $\le$ {\it T} $\le$ {\it T}$_{\rm max}$, where $a$ and $b$ are the adjustable parameters. From these views, {\it T}$_{\rm max}$ is a critical temperature, below which the spin-flip scattering manifests itself, as discussed in the previous work\cite{OBrien}. 

Hereafter, we discuss the temperature independent behavior at {\it T} $\le$ 20 K in Fig. 3. The plateau of $\Delta R_{\rm S}$ at {\it T} $\le$ 20 K indicates that the degree of spin-flip scattering is almost constant even if the temperature is lowered. In a framework of the general one-dimensional spin diffusion model\cite{VFmodel,Takahashi}, the spin signals detected by LSVs with transparent interfaces are expressed as follows\cite{Kimura1,Hamaya2}. 
\begin{equation}
\Delta R_{\rm S}  = \frac{S_{\rm N}}{S_{\rm inj}S_{\rm det}} \times \frac{\{\frac{ P_F}{(1-P_F^2)} \rho_{\rm F} \lambda_{\rm F}\}^2}{\rho_{\rm N} \lambda_{\rm N} \sinh \left( {d}/{\lambda_{\rm N}} \right)},
\end{equation}
where $S_{\rm inj}$, $S_{\rm det}$, and $S_{\rm N}$ are the areas of the junctions with a spin injector and a spin detector, the cross section of the Cu(Fe) strip, respectively, and $P_{\rm F}$ is the bulk spin polarization of the ferromagnetic electrode. $\rho_{\rm F}$ and $\lambda_{\rm F}$ are the resistivity and the spin diffusion length of the ferromagnetic electrode, and $\rho_{\rm N}$ and $\lambda_{\rm N}$ are those for the nonmagnetic wire, respectively. Since the experimentally obtained $\Delta R_{\rm S}$ is nearly constant at {\it T} $\le$ 20 K, the right-hand side of Eq. (1) should be a constant. Considering the half metallicity of the CFS spin injector and detector\cite{Wurmehl}, we assume that the values of $P_{\rm F}$ of CFS ($P_{\rm CFS}$), $\rho_{\rm CFS}$, and $\lambda_{\rm CFS}$ are constant in the low temperature regime\cite{Bombor,Hamaya2}. In addition, as the logarithmic increase in the resistivity of Cu(Fe) is suppressed in the presence of $\Delta\mu$\cite{Taniyama}, $\rho_{\rm Cu(Fe)}$ can be considered to be almost constant at {\it T} $\le$ 20 K, leading to a constant $\lambda_{\rm Cu(Fe)}$ because the relationship $\rho_{\rm Cu(Fe)} \times$ $\lambda_{\rm Cu(Fe)} \approx$ constant is fulfilled in general\cite{Dubois}. The constant $\lambda_{\rm Cu(Fe)}$ is compatible with the description that the degree of spin-flip scattering events is independent of temperature at {\it T} $\le$ 20 K and therefore the temperature regime {\it T} $<$ {\it T}$_{\rm K}$ corresponds to the low temperature Fermi liquid Kondo singlet regime. From the results, we conclude that {\it T}$_{\rm K}$ can be directly characterized by measuring the pure spin current transport in a diluted magnetic alloy. 

As we have seen before, $d$ dependence of $\Delta R_{\rm S}$ results from the change in $\Delta\mu$ in the host Cu at {\it T} $\le$ 30 K ($=$ {\it T}$_{\rm K}$). We now discuss the effect of $\Delta\mu$ on the Kondo effect in more depth. When $\Delta\mu$ is increased, the denominator of the right-hand side of Eq. (1) does not remain and $\rho_{\rm Cu(Fe)}$ and $\lambda_{\rm Cu(Fe)}$ should be both a function of $\Delta\mu$. To investigate the influence of $\Delta\mu$ on $\lambda_{\rm Cu(Fe)}$, we measured temperature dependent $\Delta R_{\rm S}$ for various spin injection currents ($I$) by using the LSV with $d$= 312 nm, as shown in Fig. 4(a). 
\begin{figure}[t]
\begin{center}
\includegraphics[width=8.5cm]{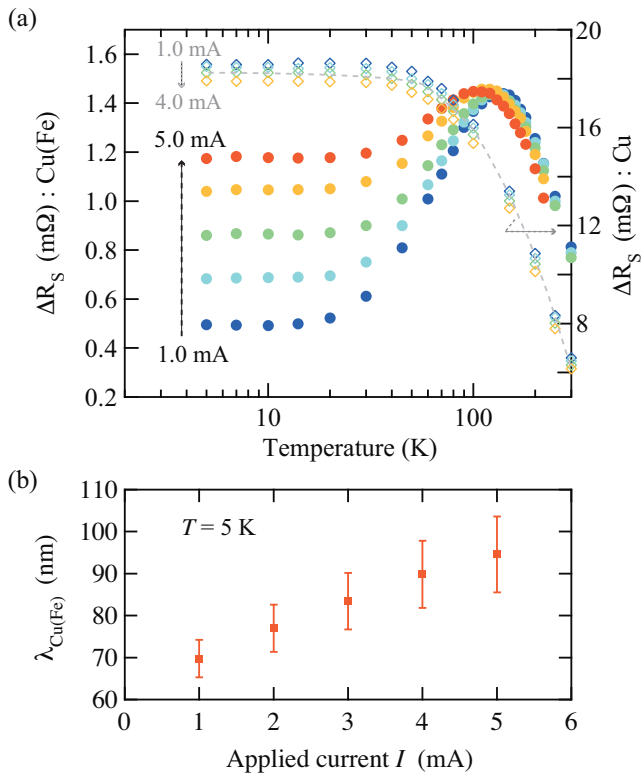}
\caption{(Color online) Temperature dependence of $\Delta R_{\rm S}$ for a CFS-Cu(Fe)-CFS LSV (left axis) with $d =$ 312 nm recorded at every 1.0 mA. The date on the right axis shows reference data for a CFS-Cu-CFS LSV with $d =$ 300 nm recorded at every 1.0 mA. (b) The estimated $\lambda_{\rm Cu(Fe)}$ as a function of $I$ at 5 K for the LSV with $d =$ 312 nm.}
\end{center}
\end{figure}  
With increasing $I$ from 1.0 to 5.0 mA, the magnitude of $\Delta R_{\rm S}$ is remarkably enhanced up to $\sim$1.2 m$\Omega$ only at {\it T} $\le$ {\it T}$_{\rm max}$ while the temperature independence of $\Delta R_{\rm S}$ at {\it T} $\le$ 30 K remains for each $I$ value. Since spin signals are generally observed to decrease with increasing $I$ due to Joule heating for conventional LSVs with a Cu channel (see right axis)\cite{Kimura1}, the enhancement in $\Delta R_{\rm S}$ is clearly an anomalous result. We also note that similar results are observed for samples with different $d$ values as presented in Fig. S1, ensuring that the effect is intrinsic under pure spin current injection conditions.

Eq. (1) now provides a rough estimate of $\lambda_{\rm Cu(Fe)}$ for various spin injection currents assuming the acceptable materials parameters; $P_{\rm CFS}\sim$ 0.6\cite{Kimura1,Hamaya2}, $\lambda_{\rm CFS}=4.0 - 5.0$ nm, $\rho_{\rm Cu(Fe)}=1.80$ $\mu\Omega$cm under the spin injection conditions at 5 K, $\rho_{\rm F}=35.1$ $\mu\Omega$cm which was obtained from our measurement. $\lambda_{\rm Cu(Fe)}$ obtained is shown in Fig. 4(b) as a function of $I$ at 5 K for the LSV with $d =$ 312 nm -- see Fig. S2 for more details. As seen in Fig. 4(b), $\lambda_{\rm Cu(Fe)}$ monotonically increases with increasing $I$, which we attribute to the suppression of the Kondo effect, viz., when $\Delta\mu$ in the host Cu becomes more significant by spin injection, the formation of Kondo clouds is suppressed due to less spin-flip scattering. 
It should be noted that a conventional way of estimating $\lambda_{\rm Cu(Fe)}$\cite{VFmodel,Takahashi} using a fit of the one-dimensional spin diffusion model to the $d$ dependence of $\Delta R_{\rm S}$ cannot be applicable any more since $\lambda_{\rm Cu(Fe)}$ is varied for different $d$ values. In fact, we have confirmed that the fitted curves largely deviate from the experimental data. This interesting phenomenon is a consequence of the generation of a large $\Delta\mu$ in the Kondo alloy by using a highly spin polarized spin injector and detector. 

We comment on the presence of the maximum in the $\Delta R_{\rm S}-T$ curves at $T_{\rm max}$. A recent study using conventional Cu-based LSVs found a similar maximum and suggested that the maximum was associated with the Kondo effect induced by Fe impurities at the Cu/ferromagnet interfaces\cite{OBrien,Casanova1,Batley,OBrien2}. Our results using Kondo alloy-based LSVs are in qualitative agreement with their claim, and the maximum is very likely related to the Kondo spin-flip scattering for the formation of Kondo clouds\cite{OBrien}. For our conventional Cu-based LSVs with CFS, however, no maximum was observed as shown on the right axis of Fig. 4(a) although CFS contains Fe element. In our view, our LSVs have the robust CFS/Cu interfaces because of the high quality epitaxial CFS electrodes grown by MBE\cite{Yamada}. The single crystalline quality of the CFS electrodes efficiently prevents the contamination of Fe in the Cu wire. A similar feature showing no maximum has been observed in our previous work\cite{Hamaya2}. Also, the fact that $\Delta R_{\rm S}$ is reduced with increasing $I$ due to Joule heating in conventional Cu-based LSVs is in clear contrast to the results for the Kondo alloy based LSVs.

In conclusion, we have demonstrated clear evidence for the suppression of the Kondo effect due to a pure spin current in LSV structures with Co$_{2}$FeSi/Cu(Fe) Kondo alloy interfaces. A remarkable reduction in the non-local spin signal has been observed associated with $s$-$d$ spin-flip scattering in the Cu(Fe) Kondo alloy. We have also shown that the non-local spin signal exhibits a plateau below the temperature regime corresponding to the formation of Fermi liquid Kondo singlets. With increasing pure spin current density, the reduction in the non-local spin signal becomes less significant, clearly indicating that $s$-$d$ spin-flip scattering associated with the Kondo effect is efficiently suppressed under greater spin accumulation conditions. From these results, we conclude that the Kondo effect can also be tuned by injecting a pure spin current into a Kondo alloy.

\vspace{10mm}
This work was partly supported by JSPS KAKENHI Grant Numbers 26103003, 252460200, 26289229, 15H01014. S. O. acknowledges JSPS Research Fellowships for Young Scientists.


\end{document}